\documentclass[showpacs,prd,superscriptaddress,onecolumn]{revtex4}

\usepackage{amsmath}
\usepackage{amssymb}
\usepackage{amsfonts}
\usepackage{graphicx,bm}
\usepackage{dcolumn}
\usepackage[colorlinks=true]{hyperref}
\usepackage{color,amsxtra}
\usepackage{epsf}
\usepackage{enumerate}
\usepackage{hhline}
\usepackage{array}
\usepackage{tabularx}
\newcommand{\be}{\begin{equation}}
\newcommand{\ee}{\end{equation}}
\newcommand{\bea}{\begin{eqnarray}}
\newcommand{\eea}{\end{eqnarray}}
\newcommand{\beaa}{\begin{eqnarray*}}
\newcommand{\eeaa}{\end{eqnarray*}}

\begin{document}

\tolerance=5000

\title{Testing viable extensions of Einstein-Gauss-Bonnet gravity}

\author{Sergei~D.~Odintsov}
\email{odintsov@ice.csic.es} \affiliation{Institute of Space Sciences (ICE, CSIC) C. Can Magrans s/n, 08193 Barcelona, Spain }
 \affiliation{Instituci\'o Catalana de Recerca i Estudis Avan\c{c}ats (ICREA),
Passeig Luis Companys, 23, 08010 Barcelona, Spain}
\author{Diego~S\'aez-Chill\'on~G\'omez}
\email{diego.saez@uva.es} \affiliation{Department of Theoretical, Atomic and Optical
Physics, Campus Miguel Delibes, \\ University of Valladolid UVA, Paseo Bel\'en, 7, 47011
Valladolid, Spain}
\author{German~S.~Sharov}
 \email{sharov.gs@tversu.ru}
 \affiliation{Tver state university, Sadovyj per. 35, 170002 Tver, Russia}
 \affiliation{International Laboratory for Theoretical Cosmology,
Tomsk State University of Control Systems and Radioelectronics (TUSUR), 634050 Tomsk,
Russia}

\begin{abstract}
Some models within the framework of Gauss-Bonnet gravities are considered in the presence of a non-minimally coupled scalar field. By imposing a particular constraint on the scalar field coupling, an extension of the called Einstein-Gauss-Bonnet gravity that keeps the correct speed of propagation for gravitational waves, is considered. The cosmological evolution for this viable class of models is studied and compared with observational data (BAO, CMB, Sne Ia,..), where we obtain the corresponding bounds for these theories and show that such theories fit well the data and provide a well-behaved cosmological evolution in comparison to the standard model of cosmology. Some statistical parameters show that the goodness of the fits are slightly better than those for $\Lambda$CDM model.


\end{abstract}

%
\maketitle
%

\section{Introduction}

Cosmology has experienced an impressive progress over the last decades, both theoretically as observationally. This has propitiated a change of paradigm in the view of cosmology, turning from a speculative science to a quantitative and precise field that can be compared with real and confident observational data. With the progression on the data measurements and on the theoretical understanding of General Relativity (GR), new conundrums have arisen that have challenged the scientific community. The modern picture of cosmology owns some uncertainties that includes the assumption of two unexplained components, the so-called dark energy and dark matter, necessary to explain the current observations. While dark matter is required to explain different astrophysical phenomena as well as the formation of large scale structure along the cosmological evolution, dark energy essentially pushes the expansion faster by a negative pressure. In most of the models, dark energy dominates over the late-time cosmological epoch, so one can say that its effects on the cosmological history are quite recent (for a review, see  \cite{Bamba:2012cp,Huterer:2017buf}). The simplest and more accepted cosmological model is the so-called $\Lambda$CDM model, which describes dark matter as a non-relativistic fluid and dark energy as a cosmological constant which effectively leads to a negative pressure. According to the recent Planck $\Lambda$CDM-based estimations, dark energy complies around seventy per cent of the total energy balance in the current universe  \cite{Planck2018}. Nevertheless, the nature of dark energy remains unclear despite the many well-motivated analysis  behind this issue \cite{Bamba:2012cp,Huterer:2017buf}. One of the most popular ways for explaining dark energy lies on frameworks that go beyond GR, suggesting that GR is not a complete theory (also supported by its well-known UV-incompleteness) so that late-time acceleration expansion is just an effect of some modifications of the underlying theory which is the real actor at large scales during the dark energy dominated epoch (for a review see \cite{Nojiri:2017ncd}). Moreover, all this activity is also motivated by the recent discrepancy among different measurements of the Hubble constant, what has been called Hubble tension. This tension arises mainly due to the differences between $\Lambda$CDM-based Planck  estimations \cite{Planck2018} and measurements by SH0ES  group \cite{Riess2021}, which implies a more than $4\sigma$ tension that requires new modified cosmological scenarios  for its explanation \cite{DiValentino:2021izs}. \\

Theories beyond GR have been widely proposed and analysed in the framework of cosmology \cite{Nojiri:2017ncd}, but one of  the most promising lines of research includes models containing the Gauss-Bonnet (GB) invariant \cite{RizosT:1994,Kawai:1998ab,Kanti:1998jd,AlexeyevTU:2000,Nojiri:2005vv,Nojiri:2006je,Cognola:2006,
Guo:2009uk,Guo:2010jr,Jiang:2013gza,KohLLT:2014,Kanti:2015pda,Yi:2018gse,Odintsov:2018zhw,NojiriOOCP:2019,Odintsov:2020sqy,
Odintsov:2020zkl,Odintsov:2020mkz,OikonomouF:2020nm,Fomin:2020,Pozdeeva:2020,PozdeevaGSTV:2020,VernovP:2021,OdintsovOF:2021,NojiriOP:2022, Nojiri:2005jg,CognolaENOZ:2006,BambaOSZ:2010,MyrzakulovST:2011,Cruz-DombrizS:2012,BenettiSCAL:2018,LeeT:2020,NavoE:2020,OdintsovOFF:2020,DeLaurentis:2015fea,Oikonomou:2017,NojiriOOP:2021}. Motivations beyond the mere cosmological aspects are also related to the presence of quadratic GB terms in low energy effective action of string theory. Hence, in order to include the GB term in the gravitational action one should note that the GB term is a topological invariant in four dimensions, such that any linear term in the action turns out a total derivative and does not introduce any modifications at the level of the field equations. Some of the several ways that GB induces real corrections are: (a) theories including scalar fields  coupled to the GB term
 \cite{RizosT:1994,Kawai:1998ab,Kanti:1998jd,AlexeyevTU:2000,Nojiri:2005vv,Nojiri:2006je,Cognola:2006,
Guo:2009uk,Guo:2010jr,Jiang:2013gza,KohLLT:2014,Kanti:2015pda,Yi:2018gse,Odintsov:2018zhw,NojiriOOCP:2019,Odintsov:2020sqy,
Odintsov:2020zkl,Odintsov:2020mkz,OikonomouF:2020nm,Fomin:2020,Pozdeeva:2020,PozdeevaGSTV:2020,VernovP:2021,OdintsovOF:2021,NojiriOP:2022}, (b) gravities with non-trivial Lagrangians (modified GB gravities)
\cite{Nojiri:2005jg,CognolaENOZ:2006,BambaOSZ:2010,MyrzakulovST:2011,Cruz-DombrizS:2012,BenettiSCAL:2018,LeeT:2020,NavoE:2020,OdintsovOFF:2020,DeLaurentis:2015fea,Oikonomou:2017,NojiriOOP:2021} and (c) models with extra spatial dimensions
\cite{ElizaldeMOOF:2007,PavluchenkoT:2009,ChirkovGT:2021} and dimensional regularization \cite{Glavan:2019inb,Fernandes:2022zrq,Fernandes:2020nbq}. Within the latter, variations in arbitrary dimensions of the gravitational action may lead to Gauss-Bonnet contributions in $D=4$ by redefining the Gauss-Bonnet coupling constant in order to remove a factor of $D-4$, calling this type of action as Einstein-Gauss-Bonnet (EGB) theory \cite{Glavan:2019inb}. Nevertheless, some papers have argued that such a limit is ill-defined and is only valid  in some particular spacetimes \cite{Arrechea:2020gjw,Cao:2021nng,Gurses:2020ofy}. On the other hand, models that present the scalar field coupled to the GB term and also to the Ricci scalar linearly, a kind of extension of Brans-Dicke theory that is can be casted as an extension of Einstein-Gauss-Bonnet (EGB) theory or scalar EGB theory \cite{Nojiri:2005vv,Nojiri:2006je,Cognola:2006,Guo:2009uk,Guo:2010jr,Jiang:2013gza,KohLLT:2014,Kanti:2015pda,
Yi:2018gse,Odintsov:2018zhw,NojiriOOCP:2019,Odintsov:2020sqy,
Odintsov:2020zkl,Odintsov:2020mkz,OikonomouF:2020nm,Fomin:2020,Pozdeeva:2020,PozdeevaGSTV:2020}. These models can potentially describe correctly both the early and late-time era. In particular, inflationary scenarios have been developed in scalar EGB gravities in previous literature \cite{Guo:2009uk,Guo:2010jr,Jiang:2013gza,KohLLT:2014,Kanti:2015pda,Yi:2018gse,Odintsov:2018zhw,NojiriOOCP:2019,Odintsov:2020sqy,
Odintsov:2020zkl,Odintsov:2020mkz,OikonomouF:2020nm,Fomin:2020,Pozdeeva:2020,PozdeevaGSTV:2020}
as well as with non-trivial Lagrangians \cite{DeLaurentis:2015fea,Oikonomou:2017}. Also scenarios with bouncing solutions have been proposed within GB gravities \cite{NojiriOP:2022,NavoE:2020,OdintsovOFF:2020}. In addition, (E)GB gravities have drawn a lot of attention within strong-field regimes, since might provide a different and richer spectrum of solutions for compact objects than in GR \cite{Rubiera-Garcia:2015yga,Tangphati:2021tcy,Hansraj:2020rvc,Singh:2020xju}, which consequently may lead to some astrophysical signs that would be used in the future to constrain gravitational theories and test departures from GR \cite{Javed:2022kzf,Suzuki:2022snr,Papnoi:2021rvw,Huang:2021bdm,Charmousis:2021npl,Shaymatov:2020yte,Blazquez-Salcedo:2020caw,Kumar:2020sag,Malafarina:2020pvl,EslamPanah:2020hoj,Zhang:2020sjh}.\\
 
Some  modified GB models of the late-time cosmological evolution have been tested with observational
data. In particular, in Ref.~\cite{BenettiSCAL:2018} a model with a non trivial function of the GB term and the Ricci scalar was tested with data including Supernovae Ia (SNe Ia) from the Joint Light-Curve Analysis, Planck (2015) Cosmic Microwave Background radiation (CMB) data and local measurements of the Hubble parameter $H_0$. A more simplified version was also confronted with the observational data in  \cite{LeeT:2020}. In both cases, the goodness of the fits suggests that this type of modified gravities can be considered as a competitive alternative for dark energy.  \\

In this paper we consider a particular EGB model that has been shown up to keep the speed of propagation of gravitational waves as the speed of light, an important constraint imposed by the analysis of the GW170817 event by the LIGO/VIRGO collaboration \cite{LIGOScientific:2017vwq}. Such constraint is not satisfied unless the coupling among the scalar field and the GB term is restricted by a particular relation \cite{Odintsov:2020sqy,Odintsov:2020zkl,Odintsov:2020mkz}. Then, we firstly analyse some previous EGB models proposed in Ref.~\cite{OdintsovOF:2021} and then concentrate on the most successful one and by using different datasets that include Pantheon SNe Ia sample  \cite{Scolnic:2017},  CMB data, BAO data and measurements of the Hubble parameter $H(z)$ (Cosmic Chronometers), we show that a particular EGB model fits as good as $\Lambda$CDM model the observational data. \\

The paper is organized as follows: section \ref{Models} reviews the basics of EGB gravity and introduce the models analysed in the paper. In sect.~\ref{Data}, the observational data including SNe Ia, BAO, CMB and $H(z)$ data is shown up as well as the procedure for fitting the model. Sect.~\ref{Results} is devoted to the results and the discussion about the implications of the fits  and its comparison with $\Lambda$CDM model. Finally, section \ref{Con} gathers the conclusions of the paper.

\section{Gauss-Bonnet gravity}
\label{Models}

\begin{figure}[th]
\centerline{ \includegraphics[scale=0.66,trim=5mm 0mm 2mm -1mm]{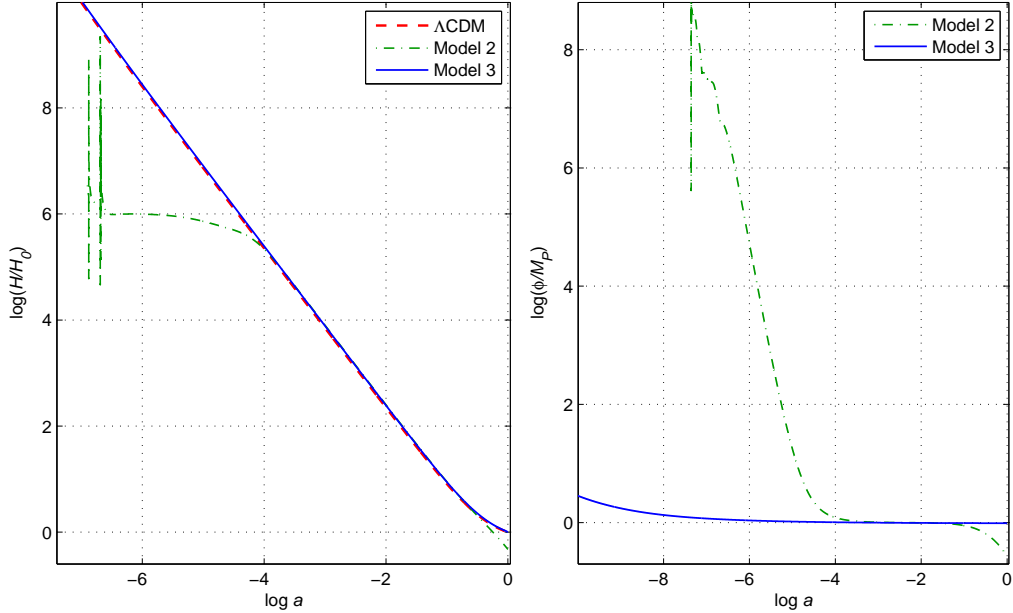}}
\caption{Evolution of the Hubble parameter $E=H/{H_0}$ (left panel) and the
scalar field $\varphi=\kappa\phi$ versus $x=\log a$ for the $\Lambda$CDM  model and the EGB
models 2 (\ref{Mod2}) and 3 (\ref{Mod3}) (parameters  given in Table~\ref{Estim}). }
  \label{F1}
\end{figure}

The central gravitational action analysed along this paper was firstly studied in Ref.~\cite{OdintsovOF:2021} and can be cast as follows:
\begin{equation}
 \label{action}
S=\int{d^4x\sqrt{-g}\left(\frac{f(R,\phi)}{2\kappa^2}-\frac{1}{2}g^{\mu\nu}\partial_\mu\phi\partial_\nu\phi-V(\phi)-\xi(\phi)\,
 \mathcal{G}+\mathcal{L}_{(m)}\right)}\ .
\end{equation}
Here $\kappa^2=8\pi G$, $\phi$ is a scalar field, $R$ is the usual Ricci scalar and $\mathcal{G}$ is the Gauss-Bonnet topological invariant that is given by:
 \begin{equation}
  \mathcal{G}=R^2-4R_{\mu\nu}R^{\mu\nu}+R_{\mu\nu\sigma\rho}R^{\mu\nu\sigma\rho}\ ,
   \label{GB}
\end{equation}
  where $R=R_\mu^\mu$, $R_{\mu\nu}$ and $R_{\mu\nu\sigma\rho}$  denote
the Ricci scalar, Ricci tensor and the Riemann curvature tensor respectively. We shall assume a flat Friedman-Robertson-Walker
(FRW) metric
 \begin{equation}
\label{metric}
ds^2=-dt^2+a(t)^2\delta_{ij}dx^idx^j\ ,
 \end{equation} 
where $a(t)$ is the scale factor. The corresponding expressions for the Ricci scalar and the Gauss-Bonnet term can be easily obtained for this metric, leading to:
\be R=6(2H^2+\dot H),\qquad \mathcal{G}=24H^2(H^2+\dot H)\ .
\ee
By varying the action (\ref{action}) with respect to the metric and to the scalar field, the FLRW equations yield:
\begin{eqnarray}
3f_1H^2&=&\kappa^2\bigg(\rho+\frac{1}{2}\dot\phi^2+V+24\dot\xi H^3\bigg)-3H\dot f_1\, ,
\label{eq1} \\
-2f_1\dot H&=&\kappa^2\bigg(\rho+P+\dot\phi^2-16\dot\xi H\dot H\bigg)+\ddot f_1-H\dot f_1\, ,
\label{eq2}\\
\ddot\phi&+&3H\dot\phi+V'(\phi)-\frac{f_1^\prime(\phi)}{2\kappa^2}R+\xi'(\phi)\,\mathcal{G}=0\, .
\label{eqphi}
\end{eqnarray}
Here $\rho$ and $P$ are the matter density and pressure respectively and we have assumed $f(R,\phi)=R\cdot f_1(\phi)$, since the main analysis of this paper is devoted to such a type of gravitational action, which is sometimes called Einstein-Gauss-Bonnet (EGB) theories due to the linearity of the function $f(R,\phi)$ with respect to the Ricci scalar. Nevertheless, some other variants of the action (\ref{action}) have been previously considered in the literature. In particular, the following ansatz was considered in Ref.~ \cite{OdintsovOF:2021}:
\be
f(R,\phi)\equiv f(R) = R + \alpha R^2 +\gamma R^\delta\ , \quad  \xi=\exp(\phi/M_{Pl})\ , \quad V=0\ ,
\ee
The analysis in \cite{OdintsovOF:2021} of the statefinder parameters for this model agrees with the values predicted by observational data. In addition, it was pointed to a subdominant role for the Gauss-Bonnet term during late-time evolution and an oscillatory behaviour for redshifts $z>5$. Another model also analysed in Ref.~\cite{OdintsovOF:2021} is described by:
\begin{equation}
 V(\phi)=V_0\bigg(\frac\phi{M_P}\bigg)^4,\qquad \xi(\phi)=\xi_0\bigg(\frac\phi{M_P}\bigg)^2,\qquad
 f(R) = R\,,
 \label{Mod2}\end{equation}
While this scenario was compared with $\Lambda$CDM model and the corresponding statefinder parameters, showing a qualitative behaviour similar to  $\Lambda$CDM model for $z<10$,  at larger redshifts $z$ the model deviates largely from the $\Lambda$CDM behaviour. For illustrative purposes, we integrate numerically the system of equations (\ref{eq1})\,--\,(\ref{eqphi}) along the variable $x=\log a$ for the model (\ref{Mod2}) and depict the evolution of $H(a)/H_0$ and $\varphi(a)=\phi/M_P$ in Fig.~\ref{F1}, where we have assumed $z_\mathrm{ini}=10$, $2\xi_0\kappa^2H_0^2=10^{-6}$, $\Omega_m^0=0.3$, $\Omega_V=0.7$ (see the definition of the parameters below (\ref{dimen})). One can infer that the model (\ref{Mod2}) reproduces  $\Lambda$CDM  model for redshifts $1<z<50$ but deviates at larger redshifts where singularities arise. These singularities appear inevitably with any initial dataset, what rules out the model (\ref{Mod2}) for describing the cosmological evolution at large redshifts.\\

Nevertheless, let us consider a model for the action (\ref{action}) where $f(R,\phi)=R\cdot f_1(\phi)$, while the coupling function $\xi(\phi)$ satisfies the following differential equation \cite{Odintsov:2020sqy,Odintsov:2020zkl,Odintsov:2020mkz,OdintsovOF:2021}:
 \begin{equation}
 \ddot\xi=H\dot\xi\, .
 \label{ddxi}  \end{equation}
 This equation can be easily integrated to be expressed in terms of the scale factor as $ \dot\xi=C\cdot a$, where $ C={}$const. In addition, the potential and the function $f_1$ are assumed to be proportional to $\phi^{-1}$, leading to \cite{OdintsovOF:2021}:
 \begin{equation}
  f_1(\phi)=\frac{\phi_0}{\phi},\qquad  V(\phi)=V_0\frac{\phi_0}{\phi},\qquad\
  \dot\xi=C a\,.
 \label{Mod3}
 \end{equation}
This is the central model to be analysed and tested in this paper, since it provides a good behaviour at every redshift. The way the model is constructed is not trivial or random but is based on important physical assumptions \cite{Odintsov:2020sqy,Odintsov:2020zkl,Odintsov:2020mkz}. In particular, the constraint equation (\ref{ddxi}) implies that the speed of propagation for gravitational waves remains the speed of light, in agreement with the observations of the GW170817 event \cite{LIGOScientific:2017vwq}. In addition, the statefinder parameter analysis points to a similar behaviour of the EGB model in comparison to $\Lambda$CDM model with the similar assumptions as given in (\ref{Mod3}) at least for late-time cosmology \cite{OdintsovOF:2021}. In order to test the EGB model (\ref{Mod3}) we have to solve the system of equations (\ref{eq1})\,--\,(\ref{eqphi}) which require to provide the complete description of the matter content $\rho=\rho_m+\rho_r$, where $\rho_m$ refers to non-relativistic particles (baryons and cold dark matter $\rho_m\simeq\rho_b+\rho_c$) and $\rho_r$ to relativistic particles (photons and neutrinos). Each matter component satisfies the usual continuity equation:
 \begin{equation}
 \dot\rho_i+3H(\rho_i+P_i)=0\,.
 \end{equation}
which can be solved for a constant equation of state parameter, leading to:
\begin{equation}
\rho=\rho_{m}^0\left(a^{-3}+X_ra^{-4}\right)\, ,
 \label{rho}\end{equation}
where $X_r=\rho_{r}^0/\rho_{m}^0$, the index $0$ denotes present day values (at $t=t_0$) and $a(t_0)=1$. Below we reduce the number of free model
parameters and do not consider $\Omega_r^0$ as an independent parameter but the ratio among radiation and cold matter is fixed as provided by Planck
\cite{Planck:2013,Odintsov:2018qug,Odintsov:2020voa}:
\begin{equation}
X_r=\frac{\rho_r^0}{\rho_m^0}=2.9656\times 10^{-4}\,. \label{Xrm}
\end{equation}
For simplicity, we define the following dimensionless variables and parameters:
 \begin{equation}
 E=\frac H{H_0},\quad\; \varphi=\kappa\phi=\frac{\phi}{M_P},\quad\; \psi=\frac{\kappa}{H_0}\dot\phi;
 \qquad\Omega_m^0=\frac{\kappa^2\rho_m^0}{3H_0^2}, \quad\;\Omega_V=\frac{\kappa^2V_0}{3H_0^2} ,
 \quad\;\lambda=\kappa^2H_0 C\,.
 \label{dimen}
 \end{equation}
 In this notation the model (\ref{Mod3}) takes the form:
 \begin{equation}
  f_1(\varphi)=\frac{\varphi_0}{\varphi},\qquad  V(\varphi)=V_0\frac{\varphi_0}{\varphi},\qquad\
  \dot\xi=\frac{\lambda}{\kappa^2H_0} a\ ,
 \label{Mod33}
 \end{equation}
 while the system of equations (\ref{eq1})\,--\,(\ref{eqphi}) can be expressed as:
 \begin{eqnarray}
f_1E^2-8\lambda aE^3&=&\Omega_m^0\big(a^{-3}+X_ra^{-4}\big)+\tfrac{1}{6}\psi^2+\Omega_Vf_1+Ef_1\psi/\varphi\, ,
\label{eq12} \\
2E\frac{dE}{dx}\big(f_1-8\lambda aE\big)&=&-\Omega_m^0\big(3a^{-3}+4X_ra^{-4}\big)-\psi^2+
\frac{\varphi_0}{\varphi^2}\bigg(E\frac{d\psi}{dx}-E\psi-2\frac{\psi^2}{\varphi} \bigg),
\label{eq22}\\
\frac{d\psi}{dx}+3\psi&+&3\frac{\Omega_V\varphi_0}{E\varphi^2}-24\frac{\lambda a}{\psi}
E^2\bigg(E+\frac{dE}{dx}\bigg)-3\frac{\varphi_0}{\varphi^2}\bigg(2E+\frac{dE}{dx}\bigg)=0\, .
\label{eqphi2}
\end{eqnarray}
Here $x=\log a$, $\frac d{dt}=H\frac d{dx}$. Then, the system  (\ref{eq12})\,--\,(\ref{eqphi2}) can be solved numerically. To do so, we fix the initial conditions at the present time ($x=0$ or $a=1$), where the following free model parameters should be specified:
 \begin{equation}
 \Omega^0_m,\quad \Omega_V,\quad H_0,\quad\lambda,\quad\varphi_0\, .
\label{5param}\end{equation}
The initial value for $E$ is naturally $E\big|_{x=0}=H_0/H_0=1$ and $\psi_0$ is determined by Eq.~(\ref{eq12}):
 \be
\psi\big|_{x=0}=-3\varphi_0^{-1}+\sqrt{9\varphi_0^{-2}-6\big[\Omega_m^0(1+X_r)+\Omega_V-1+8\lambda\big]}\ .
 \ee

Fig~\ref{F1} illustrates the evolution of the normalised Hubble parameter and the scalar field for this model in comparison to $\Lambda$CDM model and to the previous model (\ref{Mod2}). One can see that the model (\ref{Mod3}) behaves closely to $\Lambda$CDM model, avoids singularities and can be promoted to be tested with observational data. In the following sections, we test and compare the model (\ref{Mod3}) with several sources of observational data.

\section{Observational tests} \label{Data}

For testing the EGB model given in (\ref{Mod3}) with observational data, we are using observable parameters of Type Ia Supernovae (SNe Ia), Baryon Acoustic
Oscillations (BAO), Cosmic Microwave Background radiation (CMB) and estimations of the Hubble parameter $H(z)$ for a set of different redshifts $z$. For the fittings we will follow some
methods developed previously in the literature \cite{Odintsov:2017qif,Odintsov:2018qug,Odintsov:2020voa,NojiriOGS:2021,OdintsovGS:2021}.\\

The Pantheon catalogue \cite{Scolnic:2017} for SNe Ia data  is used, which includes
$N_{\mathrm{SN}}=1048$ SNe Ia data points with redshifts $0< z_i\le2.26$ and the distance moduli $\mu_i^\mathrm{obs}$. The corresponding theoretical values of the luminosity distances $D_L(z; \theta_1,\theta_2,\dots)$ and the distance modulus $\mu_i^\mathrm{th}$ are obtained as follows: 
\begin{equation}
\mu^\mathrm{th} \left( z;\theta_k \right) = 5 \log_{10} \frac{D_L \left( z;\theta_k
\right)}{10\, \mbox{pc}}\, , \quad  D_L \left( z; \theta_k \right) = (1+z)\,D_M = c (1+z) \int_0^z
\frac{d\tilde z}{H \left( \tilde z; \theta_k \right)}\ , \label{muD}
\end{equation}
where $\theta_k$ is the set of free parameters (\ref{5param}). The corresponding $\chi^2$ function is obtained as usual:
\begin{equation}
\chi^2_{\mathrm{SN}} \left( \Omega_m^0,\theta_k \right) =\min\limits_{H_0}
\sum_{i,j=1}^{1048} \Delta\mu_i\left(C_{\mathrm{SN}}^{-1}\right)_{ij} \Delta\mu_j\,
,\quad \Delta\mu_i=\mu^\mathrm{th} \left(z_i;\theta_k \right)-\mu^\mathrm{obs}_i\, ,
\label{chiSN}
\end{equation}
 where $C_{\mathrm{SN}}$ is the the covariance matrix. The Hubble constant is marginalised by minimizing the $\chi^2$ function (\ref{chiSN}), since $H_0$ can not be determined from the
data \cite{Scolnic:2017}.

Hence, by fixing a particular set of the model parameters (\ref{5param}), the system of
equations (\ref{eq12})\,--\,(\ref{eqphi2}) is solved, obtaining the Hubble parameter such that the corresponding  distance
modulus (\ref{muD}) is determined and finally the $\chi^2$ function (\ref{chiSN}) is calculated.\\


For the Baryon Acoustic Oscillations (BAO) data we consider the following  two magnitudes \cite{Eisenstein:2005}
\begin{equation}
d_z(z)= \frac{r_s(z_d)}{D_V(z)}\, ,\quad
A(z) = \frac{H_0\sqrt{\Omega_m^0}}{cz}D_V(z)\, ,
\label{dzAz}
\end{equation}
where $D_V(z)$  and the comoving sound $r_s(z_d)$ are given by:
  \begin{equation}
D_V(z)=\big[{cz D_M^2(z)}/{H(z)} \big]^{1/3}\ , \quad r_s(z)=  \int_z^{\infty}
\frac{c_s(\tilde z)}{H (\tilde z)}\,d\tilde z=\frac1{\sqrt{3}}\int_0^{1/(1+z)}\frac{da}
 {a^2H(a)\sqrt{1+\big[3\Omega_b^0/(4\Omega_\gamma^0)\big]a}}\ ,
  \label{rs2}\end{equation}
with $z_d$ being the redshift at the end of the baryon drag era. Here we use 17 BAO data points for $d_z(z)$ and 7 data points
for $A(z)$ as given in Refs.~\cite{Percival:2009xn,Blake:2011en,Beutler:2011hx,Padmanabhan:2012hf,Chuang:2012qt,
Chuang:2013hya,Ross:2014qpa,Anderson:2013zyy,Oka:2013cba,Font-Ribera:2013wce,
Delubac:2014aqe}, which are also gathered all together in some previous papers \cite{Odintsov:2017qif,Odintsov:2018qug,Odintsov:2020voa}. In addition, we assume the estimations for the ratio of baryons and photons (\ref{Xrm}), $\rho_\nu=N_\mathrm{eff}(7/8)(4/11)^{4/3}\rho_\gamma$ with $N_\mathrm{eff} = 3.046$ as given by Planck 2018 data \cite{Planck2018}. The  $\chi^2$ function yields:
\begin{equation}
\chi^2_{\mathrm{BAO}}(\Omega_m^0,\theta_1,\dots)=\Delta d\cdot C_d^{-1}(\Delta d)^T +
\Delta { A}\cdot C_A^{-1}(\Delta { A})^T\, , \label{chiB}
\end{equation}
where the corresponding vectors are:
 \begin{equation}
\Delta d_i=d_z^\mathrm{obs}(z_i)-d_z^\mathrm{th}(z_i,\dots)\, , \quad \Delta
A_i=A^\mathrm{obs}(z_i)-A^\mathrm{th}(z_i,\dots)\, .
\end{equation}
Here $C_{d}$ and $C_{A}$ are the covariance matrices for the correlated BAO data \cite{Odintsov:2018qug,Odintsov:2020voa,Percival:2009xn,Blake:2011en}.\\


Moreover, in this paper we also use the Hubble parameter $H(z)$ measurements made by the method of
``cosmic chronometer'', that includes estimation of the different ages $\Delta t$ for
passive galaxies with known variations of redshifts $\Delta z$. The values $H(z)$ are obtained via the relation:
\[ 
H (z)= \frac{\dot{a}}{a} \simeq -\frac{1}{1+z}
\frac{\Delta z}{\Delta t}\,.
\]
For the present analysis $N_H=31$ data points of $H^\mathrm{obs}(z_i)$ are used as given in Refs.~\cite{Simon:2004tf,Stern:2009ep,Moresco:2012jh,Zhang:2012mp,Moresco:2015cya,
Moresco:2016mzx,Ratsimbazafy:2017vga} within the interval $0<z<2$. These measurements are
not correlated with the above mentioned BAO data points
\cite{Percival:2009xn,Blake:2011en,Beutler:2011hx,
Padmanabhan:2012hf,Chuang:2012qt,Chuang:2013hya,Ross:2014qpa,Anderson:2013zyy,
Oka:2013cba,Font-Ribera:2013wce,Delubac:2014aqe}. Then, the $\chi^2$ function for $H(z)$ fittings yields:
\begin{equation}
\chi^2_{H}=\min\limits_{H_0} \sum_{i=1}^{N_H} \left[\frac{H^\mathrm{obs}(z_i)
 -H^\mathrm{th}(z_i; \theta_k)}{\sigma_{H,i}}\right]^2 \, .
\label{chiH}
\end{equation}


Finally, the theoretical model is also fitted by using some CMB observational parameters, which are related to the photon-decoupling epoch at redshifts close to
$z_*=1089.80 \pm0.21$ and are given by \cite{Planck2018}:
\[ 
\mathbf{x}=\left(R,\ell_A,\omega_b \right)\, ,\quad
R=\sqrt{\Omega_m^0}\frac{H_0D_M(z_*)}c\, ,\quad
\ell_A=\frac{\pi D_M(z_*)}{r_s(z_*)}\, ,
\quad\omega_b=\Omega_b^0h^2
\] 
whose  estimations are  \cite{Chen:2018dbv}:
\begin{equation}
\mathbf{x}^\mathrm{Pl}=\left( R^\mathrm{Pl},\ell_A^\mathrm{Pl},\omega_b^\mathrm{Pl} \right)
=\left( 1.7428\pm0.0053,\;301.406\pm0.090,\;0.02259\pm0.00017 \right) \, .
\label{CMBpriors}
\end{equation}
The sound horizon  $r_s(z_*)$ is obtained by the integral (\ref{rs2}) with the estimation of $z_*$ given in Refs.~\cite{Chen:2018dbv,HuSugiyama95}. The reduced baryon fraction $\omega_b^0$ is considered as the nuisance parameter in the corresponding $\chi^2$ function, which is given by:
\begin{equation}
\chi^2_{\mathrm{CMB}}=\min_{\omega_b,H_0}\Delta\mathbf{x}\cdot
C_{\mathrm{CMB}}^{-1}\left( \Delta\mathbf{x} \right)^{T}\, ,\quad
\Delta \mathbf{x}=\mathbf{x}-\mathbf{x}^\mathrm{Pl}\,,
\label{chiCMB}
\end{equation}
where the covariance matrix $C_{\mathrm{CMB}}=\| \tilde C_{ij}\sigma_i\sigma_j \|$ is described in Refs.~\cite{Odintsov:2020voa} and \cite{Chen:2018dbv}. \\

Then, the free parameters for the EGB model given in (\ref{Mod3}) are fitted with all these sources of data. In the next section, the results of these fittings are discussed and compared to $\Lambda$CDM model.

\begin{figure}[bh]
\centerline{ \includegraphics[scale=0.62,trim=5mm 0mm 2mm -1mm]{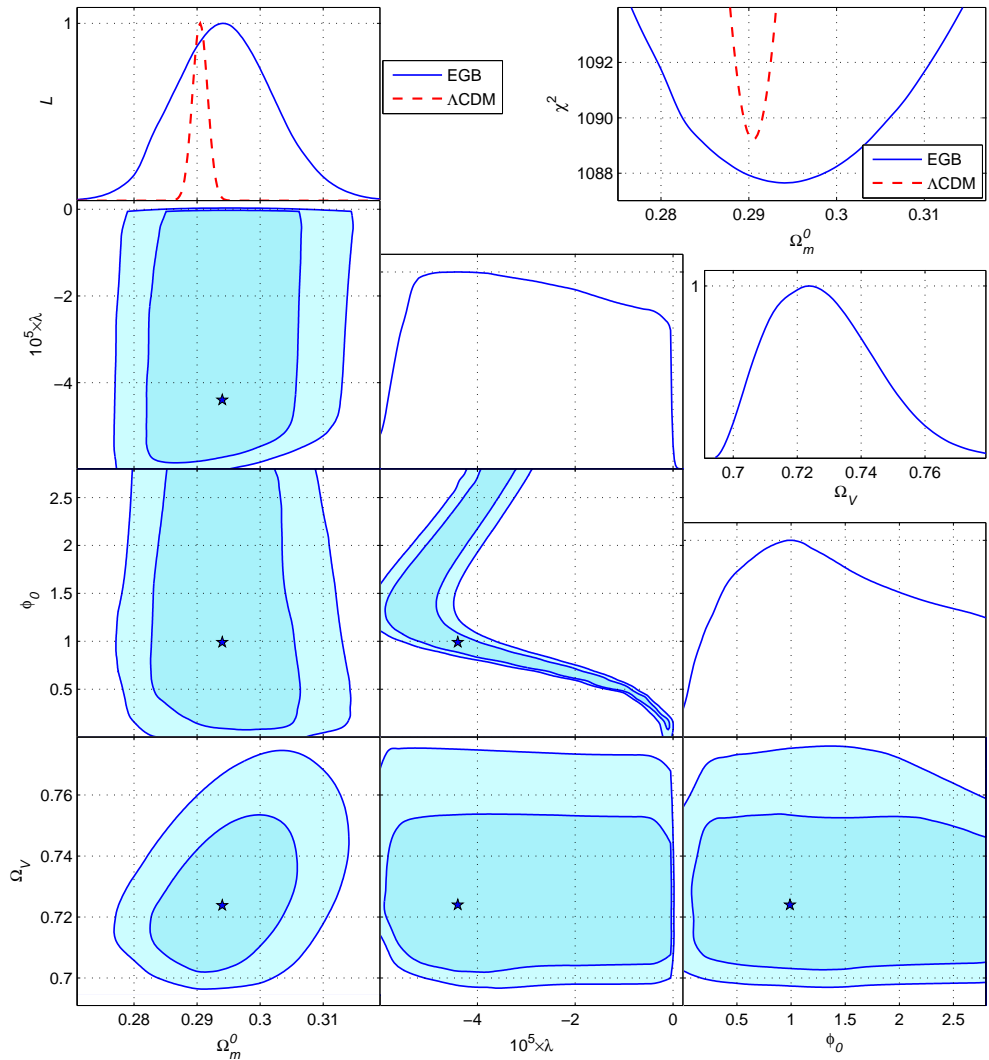}}
\caption{Contours plots, likelihoods and one-parameter distribution
$\chi^2_{\mathrm{tot}}(\Omega_m^0)$ for the EGB model (\ref{Mod3}). }
  \label{F2}
\end{figure}

\section{Results and discussion}
\label{Results}

The comparison of the model (\ref{Mod3}) with the observational datasets described above are followed by fitting its free parameters (\ref{5param}) and obtaining the corresponding $\chi^2$ functions for each dataset. The total $\chi^2_{\mathrm{tot}}$ function is the sum over all of them:
\begin{equation}
\chi^2_{\mathrm{tot}}=\chi^2_{\mathrm{SN}}+\chi^2_H+\chi^2_{\mathrm{BAO}}+\chi^2_{\mathrm{CMB}}\ .
. \label{chitot}
\end{equation}
The results are depicted in Fig.~\ref{F2} with the contour plots for each pair of the free parameters of the EGB model (\ref{Mod3}). In the panels, the blue filled contours denote the $1\sigma$ (68.27\%) and $2\sigma$ (95.45\%) confidence levels (CL) for two-parameter distributions. The corresponding minimum points for $\chi^2_{\mathrm{tot}}$ in these panels are labeled with stars. For each panel, the $\chi^2_{\mathrm{tot}}$  function is minimised over the rest of the free parameters. For instance, in the $ \Omega_m^0-\Omega_V$ panel the contours are constructed by:
\be
 \chi^2_{\mathrm{tot}}(\Omega_m^0,\Omega_V)=\min\limits_{\lambda,\varphi_0,H_0} \chi^2_{\mathrm{tot}}(\Omega_m^0,\Omega_V,\lambda,\varphi_0,,H_0)\,.
\ee
The one-parameter distribution $\chi^2_{\mathrm{tot}}(\Omega_m^0)$ in the top-right panel of Fig.~\ref{F2} is also obtained by minimizing over all the other parameters. Here we compare $\chi^2_{\mathrm{tot}}(\Omega_m^0)$  for the EGB model (\ref{Mod3}) and for flat $\Lambda$CDM model, which can be expressed as:
 \begin{equation}
H(z)=H_0\big[\Omega_m^0(a^{-3}+X_ra^{-4})+1-\Omega_m^0\big]^{1/2} .
 \label{LCDM}  \end{equation}
The best fits and  $1\sigma$ errors for the free parameters $\theta_k$ are obtained from one-parameter distributions $\chi^2_{\mathrm{tot}}(\theta_k)$ and the corresponding likelihoods (shown in Fig.~\ref{F2}):
\begin{equation}
\mathcal{L}(\theta_k)\propto \exp\left[-\frac12\chi^2_{\mathrm{tot}}(\theta_k)\right]\,.
\label{likelihood}
\end{equation}
The values of the minimum $\chi^2_{\mathrm{tot}}$ for each model together with the best fit for the free parameters are summarised in Table~\ref{Estim}. As shown, the EGB model fits better the observational data in comparison to $\Lambda$CDM model according to the minimum $\chi^2_{\mathrm{tot}}$. Nevertheless, the number of free parameters is larger for the EGB model $N_p=5$ while $\Lambda$CDM just contains $N_p=2$. For a more appropriate comparison between both models, the Akaike information criterion is followed, which relates the number of free parameters and the minimum $\chi^2_{\mathrm{tot}}$ to give a better statistical analysis through the so-called AIC parameter \cite{Odintsov:2020voa,NojiriOGS:2021,OdintsovGS:2021}:
 \begin{equation}
\mathrm{AIC} = \min\chi^2_\Sigma +2N_p
 \label{AIC} \end{equation}
As shown in Table~\ref{Estim}, such criteria favours the simpler $\Lambda$CDM model due to the shorter number of free parameters. In addition, the EGB model (\ref{Mod3}) is viable only with small negative values for the parameter $\lambda$, which leads to a nearly constant coupling with the GB term $\zeta\sim\text{constant}$, as seen in (\ref{Mod33}). Moreover, the best fit for $\varphi_0$ is close to 1 and the scalar field remains constant for redshifts $z<1000$ (see also Fig.~\ref{F1}) while the parameter  $\Omega_V$ resembles the value  $\Omega_\Lambda$ for the  $\Lambda$CDM model to some extent.\\

\begin{figure}[ht]
\centerline{ \includegraphics[scale=0.62,trim=5mm 0mm 2mm -1mm]{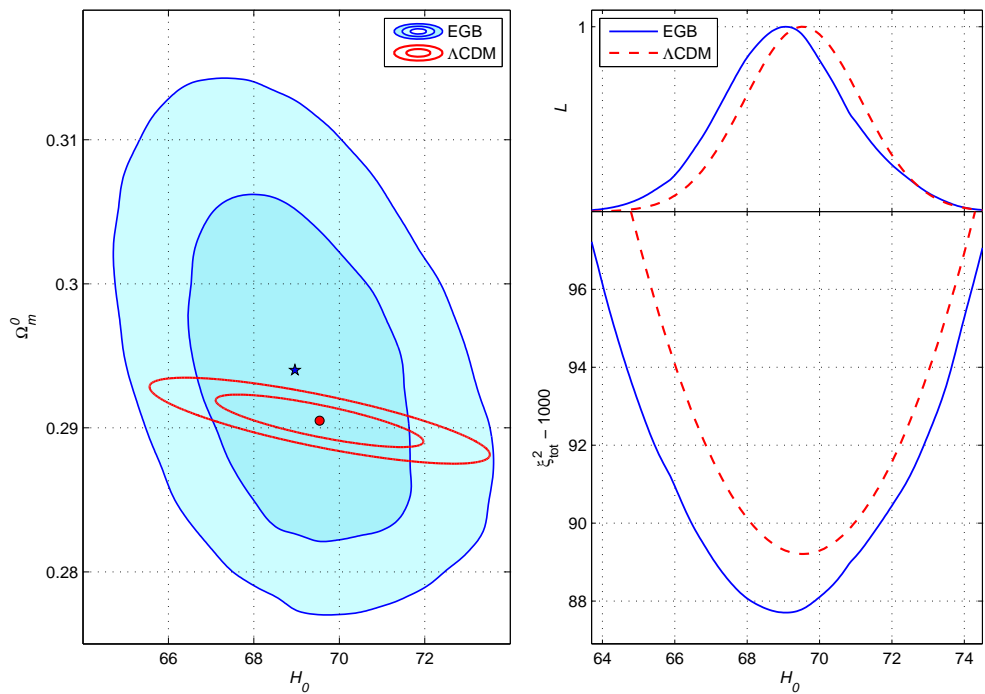}}
\caption{Contours plots, likelihoods and one-parameter distributions
$\chi^2_{\mathrm{tot}}(H_0)$ in the $H_0-\Omega_m^0$ plane for the EGB and $\Lambda$CDM
models. }
  \label{F3}
\end{figure}

In addition, Fig.~\ref{F3} shows the $H_0-\Omega_m^0$ contour plot and the distribution for the Hubble constant $H_0$ for both models. The errors are similar for the Hubble constant but larger in the case of $\Omega_m^0$ for the EGB model. The main point of these results lie on the fact that at the end this model describes effectively $\Lambda$CDM model for small redshifts while behaves well at large redshifts, providing a confident alternative to the usual cosmological model and adding more uncertainty to the dark energy paradigm.

\begin{table}[bh]
 \centering
 {\begin{tabular}{||l|c|c|c|c|c|c|c||}  \hline
 Model  & $\Omega_m^0$&  $\Omega_V$ & $H_0$ &$10^5\cdot\lambda$ & $\varphi_0$ &$ \min\chi^2_{\mathrm{tot}}\,/\,$d.o.f & AIC \rule{0pt}{1.2em}  \\
 \hline
 EGB (\ref{Mod3})& $0.294_{-0.008}^{+0.008}$ & $0.724_{-0.017}^{+0.019}$ & $68.96_{-1.62}^{+1.67}$  &
$-4.40_{-1.25}^{+4.36}$ & $0.99_{-0.73}^{+1.80}$& 1087.65 / 1102 & 1097.65 \rule{0pt}{1.2em}  \\
 \hline
$\Lambda$CDM &  $0.2905_{-0.0013}^{+0.0012}$  & - & $69.54_{-1.61}^{+1.60}$ & - & - & 1089.21 / 1105 & 1093.21 \rule{0pt}{1.2em} \\
  \hline \end{tabular}
\caption{Best fits obtained by combining the SNe Ia, $H(z)$, BAO and CMB datasets for the EGB and $\Lambda$CDM
models.}
 \label{Estim}}
\end{table}

%
\section{Conclusions}
\label{Con}

Along this paper the so-called EGB gravity has been analysed and tested with observational data. Such modifications of GR assume a particular coupling of a scalar field to the Ricci scalar and to the Gauss-Bonnet term, both of them showing up linearly in the action. Some more general actions that consider non trivial functions of the Ricci scalar have been previously considered in the literature, showing a good behaviour at late-times but inducing instabilities at large redshifts \cite{OdintsovOF:2021}. \\

Hence, we have concentrated on those gravitational actions within EGB gravity. To do so, we first analysed qualitatively the model (\ref{Mod2}) where the coupling lies just on the GB term. Despite such a model behaves well for redshifts $1<z<50$, singularities naturally arise at larger redshifts, what makes the model incompatible with a smooth and well-behaved cosmological expansion at any finite redshift. Then, the central model of the paper (\ref{Mod3}) includes also a coupling to the Ricci scalar and a particular constraint on the coupling to the GB term. This condition turns out necessary in order to keep the correct speed of propagation for gravitational waves \cite{Odintsov:2020sqy,Odintsov:2020zkl,Odintsov:2020mkz}. By analysing the cosmological evolution for this case, we found that the model keeps a regular and good behaviour at any redshifts. In addition, the model fits quite well the observational datasets used in the paper, which include the Pantheon SNe Ia data, BAO data,  $H(z)$ estimations and CMB data. The results of the fittings are summarised in Table~\ref{Estim} and depicted in Figs.~\ref{F2}, \ref{F3}, which show up similar values for the matter density $\Omega_m$ than $\Lambda$CDM model, as one would expect attending to the similar behaviour of both models. Moreover, the minimum $\chi^2$ is smaller for the EGB model which indicates a better absolute fit to the data in comparison to $\Lambda$CDM model, despite the Akaike criteria  favours the $\Lambda$CDM model, as it supports simpler models in terms of the number of free parameters. \\

Hence, we may conclude that the EGB model (\ref{Mod3}) that accomplishes some important constraints as the speed of propagation of gravitational waves, it fits quite well the corresponding cosmological data that we have at hand nowadays, specially when comparing the model with the $\Lambda$CDM model, providing a model of modified gravity that should be analysed further in order to get a better understanding of the dark energy problem and on theories beyond GR.

\begin{acknowledgments}
This work was supported in part by MINECO (Spain), project PID2019-104397GB-I00 (SDO) and PID2020-117301GA-I00 (DS-CG) and was partially supported by the program Unidad de Excelencia Maria de Maeztu CEX2020-001058-M (SDO). DS-CG is funded by the University of Valladolid (Spain) Ref. POSTDOC UVA20.
\end{acknowledgments}

\end{document}